\documentclass[aps,prx,reprint,superscriptaddress,longbibliography,amsmath,amssymb,floatfix]{revtex4-2}
\usepackage{graphicx}
\usepackage{bm}
\usepackage{capt-of}
\usepackage[colorlinks=true,linkcolor=blue,citecolor=blue,urlcolor=blue]{hyperref}

\newtheorem{theorem}{Theorem}
\newtheorem{proposition}{Proposition}
\newtheorem{lemma}{Lemma}
\newcommand{\ket}[1]{\lvert #1\rangle}
\newcommand{\bra}[1]{\langle #1\rvert}
\newcommand{\ip}[2]{\langle #1\vert #2\rangle}
\newcommand{\Tr}{\operatorname{Tr}}
\newcommand{\ran}{\operatorname{ran}}

\newcommand{\cC}{\mathcal C}
\newcommand{\cS}{\mathcal S}
\newcommand{\cH}{\mathcal H}

\newcommand{\cN}{\mathcal N}
\newcommand{\F}{\mathbb F_2}
\newcommand{\G}{\mathsf G}
\newcommand{\Glo}{P_{\rm GLOBAL}}
\newcommand{\Loc}{P_{\rm LOCC}}
\newcommand{\Sep}{P_{\rm SEP}}
\newcommand{\Lin}{P_{\triangle,\rm lin}}
\newcommand{\rc}{r_{\rm cov}}

\begin{document}

\title{Faithful filtering of constrained quantum records:\texorpdfstring{\\}{ } Local limits and collective advantage from loops}

\author{Qingtian Miao}
\email{qm8@tamu.edu}
\affiliation{Department of Physics and Astronomy, Texas A\&M University, College Station, Texas 77843, USA}
\date{November 10, 2026}

\begin{abstract}
Weak probes can record whether neighboring spins are aligned without revealing their common orientation. Reading the probes can then herald a projection of an unknown spin state onto the aligned subspace that preserves every coherence within it. We ask how often this projection succeeds when the probes are read separately or jointly. For binary pure product records and zero false positives, we derive the exact optimum over all separable measurements, and hence over all local operations and classical communication. It is the probability that one-sided local witnesses exclude every alternative, and simultaneous local measurements attain it. Joint readout can do better by accepting coherent combinations of ambiguous outcomes in which the alternatives cancel. For pairwise parity records on a connected simple graph, it beats every local strategy exactly when the graph has a loop, at all record overlaps strictly between zero and one. For general binary linear records, the condition is an irreducible dependency among at least three records. The advantage lowers the leading power of the efficiency at weak recording. On a triangle the efficiency becomes linear rather than quadratic in the witness probability, and a three-pulse coherent-light receiver captures this gain with a single click. Extensions treat targets spanning several patterns and a noisy comparison at a common output fidelity.
\end{abstract}

\maketitle

%=====================================================================
\section{Introduction}\label{sec:intro}

A weak probe can record whether two spins are aligned without revealing their common orientation. Let one such probe interact with each neighboring pair of spins in an unknown state, and let a receiver measure only the probes. When the receiver announces success, the spins can be projected onto the subspace in which all of them are aligned, with the superposition of all-up and all-down left intact. We call this heralded projection faithful filtering. Parity projections of this kind provide a route to entangling distant qubits~\cite{Barrett2005,Riste2013,Welte2021}. We ask how often faithful filtering succeeds when the same probes are read separately or jointly.

The smallest example is three spins on a triangle, with one probe per bond [Fig.~\ref{fig:scheme}]. Around the loop the three bond signs multiply to $+1$, so only four of the eight bond patterns occur. They are the all-aligned target $000$ and the alternatives $110$, $101$ and $011$, where $1$ marks an anti-aligned bond. Each pattern leaves the three probes in a product of nonorthogonal states. A local outcome that certifies one aligned bond excludes two alternatives but not the third, so a local receiver needs two such witnesses. A joint measurement can instead accept a coherent combination of ambiguous outcomes in which all three alternatives cancel and the target survives. If $s$ is the probability of a witness on an aligned bond, the best local efficiency begins at $3s^2$ at weak recording, whereas the joint efficiency begins at $4s/3$.

A gain over one particular local receiver would leave open whether a cleverer adaptive protocol could close the gap. We therefore derive the exact optimum over all separable measurements, and hence over every protocol of local operations and classical communication (LOCC), for binary pure product records and any family of alternatives. This optimum is the probability that the witnesses exclude every alternative, and simultaneous local measurements whose results are collected afterwards attain it. Against this benchmark, joint readout of pairwise $Z_uZ_v$ records on a connected simple graph gains exactly when the graph has a loop, at every record overlap $0<c_e<1$. For general binary linear records the condition is an irreducible dependency among at least three records. Whenever the gain exists, it lowers the power of $s$ with which the efficiency vanishes at weak recording.

The geometry of optimal pure-state filtering is established~\cite{SunBergouHillery2002,Bergou2003,BergouHerzogHillery2005}, as are criteria for positive LOCC discrimination~\cite{Chefles2004,DuanFengJiYing2007} and optimal local discrimination of two pure product states~\cite{ChenYang2001,JiCaoYing2005}. Larger product-state families can favor joint measurements~\cite{Bennett1999,HaKim2021,Murshid2026}. Related collective advantages occur in decoding~\cite{Mandal2026,Shutty2026,BuzetChailloux2025,TKG2013} and optical reception~\cite{Guha2011,CuiGuha2025,vanEnk2002,Sidhu2023}. Here the exact local benchmark identifies which record constraints give an intrinsic joint advantage.

The triangle also has a simple optical realization. When spins in optical cavities imprint their bond parities on coherent pulses, the three record pulses can be interfered so that every alternative leaves a monitored mode in vacuum while the target can produce a click. The amplitude cancellation then becomes a single detection event, and this fixed receiver attains the optimal weak-light order. After the single-pattern analysis, we show that targets spanning several bond patterns additionally require erasing which-pattern information, and we compare readouts at a common output fidelity in a noisy four-spin example.

Section~\ref{sec:model} defines the experiment and faithful projection. The triangle, exact local bound, loop criterion and weak-recording gain are developed in Secs.~\ref{sec:triangle}--\ref{sec:weak}, followed by the optical receiver in Sec.~\ref{sec:optics}. Sections~\ref{sec:multi} and \ref{sec:noise} treat the two extensions. Proofs and quantitative details are in Appendices~\ref{app:faithful}--\ref{app:multi}.

%=====================================================================
\section{Records and faithful filtering}\label{sec:model}

\subsection{Preparing and writing a record}

Consider $N$ spins in an unknown state $\rho$, possibly entangled with a reference system. The spins sit on the vertices of a connected graph $G$ with $m$ edges. Each edge $e=uv$ carries one probe, which records the bond parity $Z_uZ_v$. The parity is $+1$ if the two spins are aligned and $-1$ if they are anti-aligned. The probe thus learns the relative orientation of the two spins but not their common orientation.

The experimenter prepares each probe independently in
\begin{equation}
 \ket{\chi_p}=\sqrt{1-p/2}\ket0+\sqrt{p/2}\ket1,\qquad 0\le p\le1,
 \label{eq:probe-preparation}
\end{equation}
and uses it to control a parity gate, which applies $Z_uZ_v$ to the spins if the probe is in $\ket1$ and leaves them alone if it is in $\ket0$. With the spin factor written first, the gate is
\begin{equation}
 U_e=I\otimes\ket0\bra0_e+Z_uZ_v\otimes\ket1\bra1_e.
 \label{eq:controlled-parity}
\end{equation}
On a parity eigenstate, $Z_uZ_v$ multiplies the $\ket1$ component of the probe by its eigenvalue $\pm1$. The probe is therefore left in one of two records,
\begin{equation}
 \ket{a_\pm}=\sqrt{1-p/2}\ket0\ \pm\sqrt{p/2}\ket1,
 \label{eq:records-p}
\end{equation}
and the parity is stored in their relative phase. Both $\ket{\uparrow\uparrow}$ and $\ket{\downarrow\downarrow}$ leave the same record $\ket{a_+}$.

The preparation parameter $p$ sets how distinguishable the two records are. At $p=0$ they coincide and carry no parity information. At $p=1$ they are orthogonal, and one measurement reads the parity. In between, a measurement can be conclusive only part of the time. Projecting onto the unit vector $\ket{a_-^\perp}$ orthogonal to $\ket{a_-}$ in the record plane never clicks on an anti-aligned bond, so a click certifies that the bond is aligned. We call this click a \emph{witness}. The record overlap and the witness probability on an aligned bond are
\begin{equation}
 c=\ip{a_+}{a_-}=1-p,\qquad s=1-c^2=2p-p^2.
 \label{eq:strength}
\end{equation}
Weak recording means $p\to0$, so that $c\to1$ and $s\to0$. We keep $p$ for the preparation, $c$ for the record geometry and $s$ for the efficiency of witness-based readout.

Absorbing the probe preparation into the gate, $V_e=U_e(I\otimes\ket{\chi_p})$, gives the record map of one bond,
\begin{equation}
 V_e=\Pi^e_+\otimes\ket{a^{(e)}_+}+\Pi^e_-\otimes\ket{a^{(e)}_-},
 \label{eq:coupling}
\end{equation}
where $\Pi^e_\pm=(I\pm Z_uZ_v)/2$ projects the spins onto aligned or anti-aligned bond $e$. Probes prepared with different $p_e$ have overlaps $c_e=1-p_e$ and witness probabilities $s_e=1-c_e^2$. For identical probes we drop the index $e$. All strict-advantage statements below assume $0<c_e<1$, where the records are neither identical nor orthogonal.

\subsection{Legal patterns and the target}

Write $x_e=0$ for an aligned (positive) bond and $x_e=1$ for an anti-aligned (negative) one. A spin configuration fixes a bond pattern $x$, and the probes then hold the product record $\ket{w_x}$, with $\ket{a_-}$ on negative bonds and $\ket{a_+}$ elsewhere. Around any cycle the bond signs multiply to $+1$, since each spin appears in two of its bonds. A triangle therefore permits $000$, $110$, $101$ and $011$, but no pattern with an odd number of negative bonds. The graph restricts which records can occur, and this restriction is what joint readout will exploit.

Let $\cC$ be the set of legal bond patterns and $\Pi_x$ the projector onto spin configurations with pattern $x$. All probes together implement
\begin{equation}
 V=\prod_eV_e=\sum_{x\in\cC}\Pi_x\otimes\ket{w_x}.
 \label{eq:V}
\end{equation}
Each conditional record $\ket{w_x}$ is a product state, but a superposition of spin patterns generally becomes entangled with the probes.

The main task targets the all-aligned pattern $x=0$. On a connected graph its spin subspace is
$\ran\Pi_0=\operatorname{span}\{\ket{\uparrow}^{\otimes N},\ket{\downarrow}^{\otimes N}\}$.
A successful run should project the spins onto this subspace while keeping whatever superposition of all-up and all-down, and whatever correlation with the reference, the input had there. Appendix~\ref{app:faithful} extends the setting to commuting Pauli checks and reduces any other legal single-pattern target to $x=0$ by relabeling records.

\begin{figure*}[t]
\centering
\includegraphics[width=\textwidth]{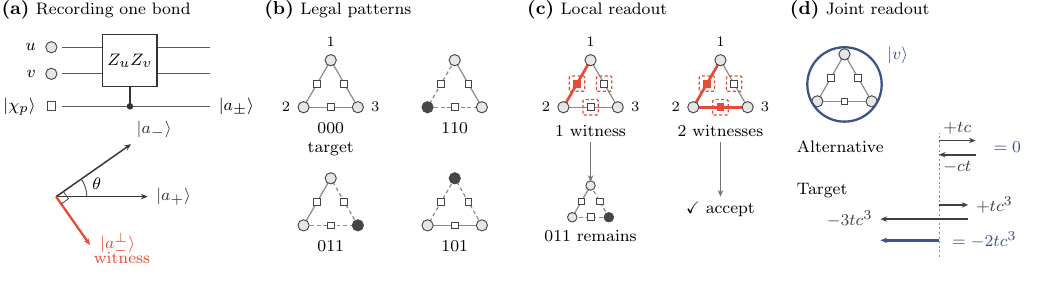}
\caption{The same recorded spins, read separately or jointly. (a) A probe prepared in $\ket{\chi_p}$ controls $Z_uZ_v$ and is left in $\ket{a_+}$ or $\ket{a_-}$ for an aligned or anti-aligned bond. The two records enclose the angle $\theta$, with $\cos\theta=c$, and the witness direction $\ket{a_-^\perp}$ is orthogonal to $\ket{a_-}$. (b) The four legal patterns of the triangle. A dark spin is flipped relative to the others, and dashed bonds are anti-aligned. (c) Separate readout measures each probe on its own (dashed boxes). A witness on bond $12$ leaves the alternative $011$, whereas witnesses on two bonds exclude all three alternatives. (d) A joint measurement accepts the direction $\ket v$ of Eq.~\eqref{eq:v}. The arrows show the contributions to the overlap of an alternative and of the target with the numerator of $\ket v$. They cancel for the alternative and leave $-2tc^3$ for the target. Both readouts herald the same spin projection and differ only in efficiency.}
\label{fig:scheme}
\end{figure*}

\subsection{Readout and successful projection}\label{sec:task}

After recording, the probes go to a receiver, who measures them and announces success or failure without acting on the spins. A local receiver is restricted to local operations and classical communication (LOCC). Each probe is then held in its own laboratory with local ancillas, and the laboratories may exchange classical messages over any number of rounds but share no entanglement. A joint receiver (GLOBAL) may instead perform any measurement on all probes together. To bound every LOCC protocol at once, we also use the larger class of separable measurements (SEP), whose effects are sums of positive product operators across probes~\cite{Bennett1999}.

Let $0\le F\le I$ be the total probe effect of all outcomes announced as success. Faithful filtering first requires that spins with an alternative pattern never be accepted, that is, zero false positives,
\begin{equation}
 \bra{w_x}F\ket{w_x}=0\quad\text{for all }x\in\cC\setminus\{0\}.
 \label{eq:zfp}
\end{equation}
For a single target pattern this condition is also sufficient. The unnormalized spin state on success is then
\begin{equation}
 \cH_F(\rho)=P_F\,\Pi_0\rho\Pi_0,\qquad P_F=\bra{w_0}F\ket{w_0}.
 \label{eq:faithful}
\end{equation}
The condition suffices because every state in the target subspace leaves the \emph{same} probe record $\ket{w_0}$. Once the alternatives are rejected, the accepted outcomes carry no information about which target component was present, so the coherence between all-up and all-down survives. This holds for every input, including one entangled with a reference (Appendix~\ref{app:faithful}).

In Eq.~\eqref{eq:faithful}, $P_F$ is the conditional efficiency, the probability of success when the spins have the target pattern. Multiplying it by the target weight gives the success probability per experimental round, $P_F\Tr(\Pi_0\rho)$. On success the spins are left in $\Pi_0\rho\Pi_0/\Tr(\Pi_0\rho)$. For example, the product input $\ket+^{\otimes N}$, with $\ket+=(\ket\uparrow+\ket\downarrow)/\sqrt2$, has target weight $2^{1-N}$ and is projected onto the Greenberger--Horne--Zeilinger state $\mathrm{GHZ}_N$. This state does not depend on which faithful receiver was used, so receivers differ only in how often they succeed. We therefore compare their best conditional efficiencies $\Glo$, $\Sep$ and $\Loc$ subject to Eq.~\eqref{eq:zfp}. The triangle shows that changing only the readout already changes this efficiency.

%=====================================================================
\section{Local and joint readout of a triangle}\label{sec:triangle}

Number the spins $1,2,3$ and list bond signs in the order $(12,23,31)$. Starting from aligned spins, flipping one spin makes its two bonds negative. Flipping spin $2$, for example, gives the pattern $110$. The legal patterns and the probe records they leave are
\begin{equation}
\begin{aligned}
 000&\mapsto\ket{a_+a_+a_+},&\quad 110&\mapsto\ket{a_-a_-a_+},\\
 011&\mapsto\ket{a_+a_-a_-},&\quad 101&\mapsto\ket{a_-a_+a_-},
\end{aligned}
\label{eq:tri-records}
\end{equation}
with $000$ the target [Fig.~\ref{fig:scheme}(b)].

\emph{Local readout.} Each probe is measured in $\{P,I-P\}$, where $P=\ket{a_-^\perp}\bra{a_-^\perp}$ is the witness projection of Sec.~\ref{sec:model} [Fig.~\ref{fig:scheme}(a)]. A witness on bond $12$ shows that this bond is aligned. It excludes $110$ and $101$ but not $011$, which also has an aligned bond $12$ [Fig.~\ref{fig:scheme}(c)]. A single witness therefore leaves one alternative open. A rule without false positives must wait for witnesses on two bonds, which together exclude all three alternatives. Measuring all three probes at once, collecting the outcomes and accepting on two or more witnesses gives
\begin{equation}
 P^{\triangle}_{\rm loc}=3s^2-2s^3 .
 \label{eq:tri-loc}
\end{equation}
Theorem~\ref{thm:local} below shows that no separable or LOCC readout does better.

\emph{Joint readout.} Outcome strings with one witness or none are individually ambiguous, and the local rule must reject them. A joint measurement on the three probes can instead accept a superposition of such strings. Write $\ket{a_+}=c\ket{a_-}+t\ket{a_-^\perp}$ with $t=\sqrt{1-c^2}$, and consider
\begin{equation}
 \ket v=\frac{t\ket{a_-a_-a_-}-c\sum_{\circlearrowleft}\ket{a_-^\perp a_-a_-}}{\sqrt{3-2s}},
 \label{eq:v}
\end{equation}
where the sum runs over the three cyclic placements of $\ket{a_-^\perp}$. An alternative, say $110$, has $\ket{a_+}$ on one probe and $\ket{a_-}$ on the other two. Its overlap with the numerator of $\ket v$ has two contributions, $tc$ from the first term and $-ct$ from the term with $\ket{a_-^\perp}$ on the aligned bond, and they cancel. The target, with $\ket{a_+}$ on every probe, gives $tc^3-3tc^3=-2tc^3\neq0$ instead. Every term of $\ket v$ has at most one witness, so $\ket v$ is orthogonal to the local acceptance projector. Adding $\ket v\bra v$ to that projector still rejects every alternative and attains the optimal joint efficiency, which follows from the pure-state filtering geometry of Ref.~\cite{Bergou2003},
\begin{equation}
 P^{\triangle}_{\rm GLOBAL}=3s^2-2s^3+\frac{4s(1-s)^3}{3-2s}=\frac{s(4-3s)}{3-2s}.
 \label{eq:tri-glo}
\end{equation}
At weak recording the joint efficiency is $4s/3$, against $3s^2$ for the best local readout. Nearly all of it comes from $\ket v$, that is, from outcomes that the local rule must discard. With coherent-light records this cancellation becomes destructive interference, which leaves one optical mode in vacuum for all three alternatives (Sec.~\ref{sec:optics}).

Separate readout can only accumulate certain evidence, so its efficiency is the probability that this evidence excludes every alternative. Joint readout gains because the three bond signs have only two independent values. The loop therefore leaves three alternatives to reject instead of seven, even though the four legal records remain linearly independent. With fewer directions to annihilate, the accepted subspace has room for vectors such as $\ket v$ that no product of witnesses reaches. Section~\ref{sec:local} turns the first observation into an exact limit for all local readouts, and Sec.~\ref{sec:class} identifies which record constraints make room for the second.

%=====================================================================
\section{The exact separable and LOCC limit}\label{sec:local}

On a general graph the triangle's two-witness rule becomes a connectivity test. Measure every probe in $\{P_e,I-P_e\}$, with $P_e=\ket{a_-^{(e)\perp}}\bra{a_-^{(e)\perp}}$, and collect the witnessed bonds. If they connect all spins, every spin is aligned with every other, and the only consistent pattern is the target. If they leave the spins in several pieces, flipping every spin of one piece changes only unwitnessed bonds. This produces an alternative that agrees with every witness. A local rule without false positives must therefore accept only when the witnessed bonds connect all spins.

When the spins have the target pattern, every probe holds $\ket{a_+}$ and the witnesses occur independently with probabilities $s_e$. The rule then succeeds with the probability that the witnessed bonds connect the graph, the all-terminal reliability~\cite{Colbourn1987}
\begin{equation}
 \Loc=\Sep=\mathcal R_G(s)=\sum_R\prod_{e\in R}s_e\prod_{e\notin R}(1-s_e),
 \label{eq:reliability}
\end{equation}
where the sum runs over connected spanning bond sets $R$. Theorem~\ref{thm:local} below shows that no separable or LOCC readout does better. Adaptive measurements and further rounds of communication thus cannot improve on simultaneous local measurements whose outcomes are collected afterwards.

The theorem does not need a graph. Describe an alternative by the nonempty set $S$ of probes that carry $\ket{a_-}$ instead of $\ket{a_+}$. A witnessed set $T$ excludes this alternative exactly when $T\cap S\ne\varnothing$.

\begin{theorem}[Local limit]\label{thm:local}
Let probe $e$ have two pure records $\ket{a^{(e)}_\pm}$ with overlap $|\ip{a^{(e)}_+}{a^{(e)}_-}|=c_e\in(0,1)$, and let $\cS$ be any family of nonempty alternative supports. The largest target acceptance over separable effects $0\le F\le I$ with zero acceptance on every alternative is
\begin{equation}
 H_\cS=\Pr\bigl[\,T\cap S\ne\varnothing\quad\forall S\in\cS\,\bigr],
 \label{eq:hitting}
\end{equation}
where each probe belongs to $T$ independently with probability $s_e=1-c_e^2$. Measuring all probes locally in $\{P_e,I-P_e\}$ at the same time, then collecting the outcomes and accepting exactly the covering sets $T$, attains this value. Hence $\Loc=\Sep=H_\cS$.
\end{theorem}

The upper bound rests on positivity. A separable effect is a sum $F=\sum_\ell\bigotimes_e A_e^{(\ell)}$ of products of positive local factors, so every term has nonnegative expectation on every record. Zero acceptance of an alternative therefore forces \emph{each} term to annihilate it. A term that contributes on the target is nonzero on $\ket{a_+}$ at every probe. It can annihilate the alternative only on a probe where the alternative carries $\ket{a_-}$, and there its local factor must lie along the witness direction. The witness set of each contributing term thus covers every alternative, and all such terms fit inside the projector onto covering outcome strings, whose target probability is $H_\cS$. Positive terms cannot cancel one another, which is why separable readout cannot use the ambiguous amplitudes that the joint triangle filter accepts. Appendix~\ref{app:local} gives the complete operator argument and a stronger version that requires only the acceptance effect to be separable.

The legal patterns form a binary linear \emph{record code} $\cC\subseteq\F^m$. For a graph this is the cut code, whose codewords are the sets of bonds cut by flipping some subset of spins. It has dimension $k=N-1$. For a general code let $\G$ be a full-row-rank $k\times m$ generator matrix with columns $\mathsf g_e$, so that $\cC=\{u^{\mathsf T}\G:u\in\F^k\}$. A witnessed set covers every nonzero codeword exactly when its columns have rank $k$. This rank condition is the algebraic form of connectivity used below. Commuting Pauli checks lead to the same linear-code setting after relabeling a legal target (Appendix~\ref{app:faithful}).

The bound also holds for optical modes and other higher-dimensional probes, because compressing each probe onto its two-dimensional record plane preserves separability and all record probabilities (Appendix~\ref{app:local}). Section~\ref{sec:optics} uses it as the exact benchmark for every per-mode optical readout.

%=====================================================================
\section{Loops make joint readout useful}\label{sec:class}

On a tree every bond pattern is legal, including those that differ from the target on a single bond. Local readout must then witness every bond, and joint readout does no better, with both efficiencies equal to $\prod_e s_e$.

A faithful effect must annihilate every alternative record and hence their span $W$. The best joint measurement therefore projects onto $W^\perp$, and its efficiency is the squared length of the target component outside $W$~\cite{Bergou2003}. On a tree $W^\perp$ contains only the string with a witness on every bond. A loop removes alternatives and enlarges $W^\perp$, which can then contain coherent superpositions of locally ambiguous strings, as $\ket v$ does on the triangle. Removing alternatives is not by itself enough, however. The following theorem identifies exactly which constraints give joint readout a strict advantage.

\begin{theorem}[Classification]\label{thm:class}
For a single target pattern of a binary linear record code $\cC\subseteq\F^m$ with generator matrix $\G$ and pure record overlaps $c_e\in(0,1)$, the following three statements are equivalent. (i) $\Glo>\Loc$. (ii) The columns of $\G$ contain a circuit, that is, a minimal linearly dependent set, with at least three elements. (iii) After deleting coordinates that vanish on all of $\cC$, the code is not a direct sum of repetition codes, each of which has only an all-zero and an all-one codeword on its block of coordinates. The criterion does not depend on the overlap values. Whenever it holds, single-probe measurements and classical control, with one joint measurement on three probes in one branch, suffice to beat $\Loc$.
\end{theorem}

For the cut code of a connected simple graph, these circuits are its cycles~\cite{Oxley2011}. A loop is therefore necessary and sufficient for a strict joint advantage at every choice $0<c_e<1$. Repeated records show why the circuit needs at least three elements. Two probes recording the same bond pose a single two-state problem with a smaller combined overlap, and local readout already attains its joint optimum~\cite{ChenYang2001,JiCaoYing2005}. A dependency among three or more records instead constrains several distinct bonds at once and opens room for the cancellation used on the triangle.

A larger loop uses the same mechanism. Choose three bonds of a cycle and measure the other bonds locally. In the branch in which a suitable set of other bonds has witnessed, those bonds tie the spins into three groups, and the three chosen bonds then carry exactly the triangle's legal patterns. A joint measurement on these three probes can accept the extra direction $\ket v$. Appendix~\ref{app:class} gives this construction for arbitrary binary linear record codes and unequal overlaps, together with the proof that repetition blocks give no gain. The three-probe construction guarantees a strict gain. The unrestricted joint optimum can gain more and can have a lower weak-recording exponent (Sec.~\ref{sec:weak}).

On the four-spin loop $C_4$ at $c=4/5$ ($p=1/5$), for example, the joint and local conditional efficiencies are about $0.27$ and $0.14$. For the input $\ket+^{\otimes4}$ both are multiplied by the same target weight $1/8$, giving success probabilities per round of about $0.033$ and $0.017$. Figure~\ref{fig:class} shows how the ratio grows as recording weakens, together with the intermediate efficiency of the three-probe construction.

%=====================================================================
\section{Weak recording: how large is the gain?}\label{sec:weak}

Let all overlaps be equal and take $s=1-c^2\to0$ at a fixed graph or code. Each local witness costs one factor of $s$. The smallest covering sets contain $k$ independent bonds or checks, so the exact local efficiency begins at order $s^k$. On a graph they are spanning trees and $k=N-1$. The triangle already shows that coherent cancellation can lower this power.

The joint power depends on how close the legal records come to linear dependence as $c\to1$. This is measured by the eigenvalues of their Gram matrix $\Gamma_{xy}=\ip{w_x}{w_y}$. Because $\Gamma_{xy}$ depends only on $x+y$, its eigenvectors are labeled by $\sigma\in\F^k$, and its eigenvalues are $2^k\pi(\sigma)$, with
\begin{equation}
 \begin{gathered}
 \pi(\sigma)=\Pr[\G Z=\sigma],\\
 \Pr(Z_e=1)=q=(1-c)/2=p/2,
 \end{gathered}
 \label{eq:pi}
\end{equation}
where the $Z_e$ are independent auxiliary bits. Here $\sigma$ labels an eigenvector of $\Gamma$, not a legal bond pattern $x$. The optimal joint efficiency is the harmonic mean of these eigenvalues (Appendix~\ref{app:global}), so the smallest ones control it. An eigenvalue whose $\sigma$ requires at least $r_\sigma$ nonzero auxiliary bits is of order $q^{r_\sigma}$. Since $q\simeq s/4$ at weak recording, the joint efficiency begins at order $s^{\rc}$, with
\begin{equation}
 \rc=\max_{\sigma\in\F^k}\ \min_{\G z=\sigma}|z|.
\end{equation}
This is the covering radius of the dependency code $\cC^\perp$, which consists of the vectors $z$ with $\G z=0$. It counts the powers of $s$ that the joint filter must pay.

\begin{proposition}[Two exponents]\label{prop:exponents}
For a fixed binary linear record code with nonzero columns, as $s\to0$,
\begin{align}
 \Loc&=b\,s^k[1+O_\cC(s)],\nonumber\\
 \Glo&=A_\cC\,s^\rc[1+O_\cC(s)],
 \label{eq:exponents}
\end{align}
where $b$ and $A_\cC$ are positive code-dependent constants specified in Appendix~\ref{app:weak}, and $O_\cC(s)$ denotes a correction bounded by a code-dependent constant times $s$. Always $\rc\le k$, with equality if and only if no circuit has three or more elements.
\end{proposition}

Together with Theorem~\ref{thm:class}, the proposition shows that the single-pattern joint advantage, whenever it exists, lowers the leading power of the recording strength. Figure~\ref{fig:class} shows the change from $s^3$ to $s^2$ on $C_4$. For bond parities on any connected graph,
\begin{equation}
 \lfloor N/2\rfloor\le\rc\le N-1.
 \label{eq:halving}
\end{equation}
Coherent readout of pairwise $Z_uZ_v$ records, or of records equivalent to them under local basis changes, can therefore at most halve the local exponent. Odd cycles reach exactly half. Even cycles stay above half and approach it only as their length grows. The lower bound follows from the relation between the cycle-code covering radius and minimal $T$-joins~\cite{SoleZaslavsky1993}. For an even set $T$ of vertices, a $T$-join is a set of edges whose odd-degree vertices are exactly those in $T$~\cite{EdmondsJohnson1973}. Appendix~\ref{app:weak} gives the counting argument.

These exponents describe the conditional efficiency at fixed code size, and Appendix~\ref{app:weak} gives their prefactors. Heralding a GHZ state from $\ket+^{\otimes N}$ multiplies the local and joint efficiencies by the same target weight $2^{1-N}$. On the triangle the joint advantage lowers the power from $s^2$ to $s$, and the next section shows that a simple optical receiver captures this first-order gain.

\par\medskip
\noindent\begin{minipage}{\columnwidth}
\centering
\includegraphics[width=\columnwidth]{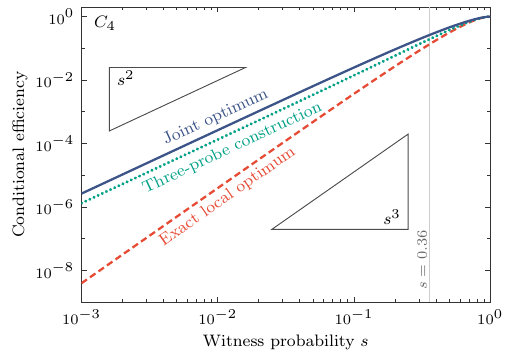}
\captionof{figure}{Conditional efficiencies on the four-spin loop $C_4$ for identical ideal records and zero false positives. The joint optimum (solid) and the three-probe construction of Sec.~\ref{sec:class} (dotted) scale as $s^2$, whereas the exact local optimum $\Sep=\Loc$ (dashed) scales as $s^3$. The vertical line marks $s=0.36$, the example $c=4/5$ of Sec.~\ref{sec:class}. The curves follow the closed forms of Appendices~\ref{app:global} and \ref{app:class}.}
\label{fig:class}
\end{minipage}
\par\medskip

%=====================================================================
\section{A coherent-light receiver}\label{sec:optics}

\subsection{Writing optical records}

\begin{figure*}[t]
\centering
\includegraphics[width=\textwidth]{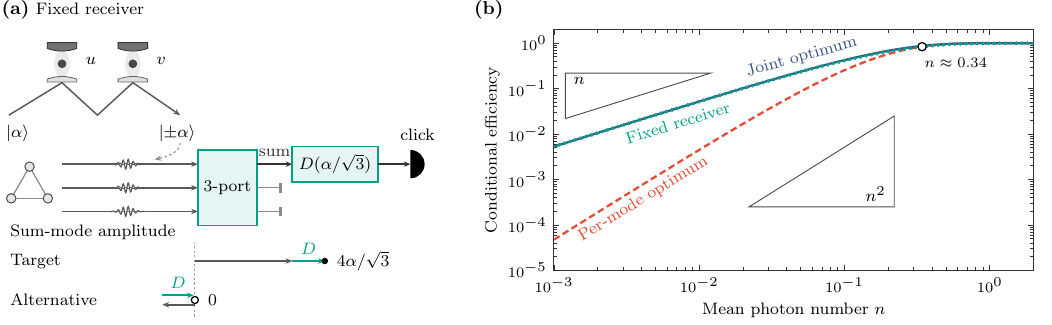}
\caption{Coherent-light receiver for the triangle. (a) Each record pulse reflects from the cavities of its two spins and leaves as $\ket{\pm\alpha}$. The three pulses (wave packets) enter a balanced three-port together, and the sum mode is displaced by $D(\alpha/\sqrt3)$ and detected, while the other two outputs are discarded. The lower diagram shows the sum-mode amplitude and its shift by the displacement $D$. Every alternative is displaced to vacuum, so a click has no false positives, whereas the target amplitude becomes $4\alpha/\sqrt3$. (b) Conditional efficiencies versus the mean photon number $n$ per pulse, for ideal records and zero false positives. The fixed receiver (dotted) follows the joint optimum (solid) at weak light, and both scale as $n$. The exact optimum of all per-mode readouts (dashed) scales as $n^2$. The open circle marks $n\simeq0.34$, beyond which the per-mode optimum overtakes the fixed receiver. Heralding probabilities are these efficiencies times the target weight, for example $1/4$ for $\ket+^{\otimes3}$.}
\label{fig:optics}
\end{figure*}

Coherent light gives the triangle cancellation a direct optical form. Let each spin be an atom in a one-sided optical cavity, as in atom--cavity reflection gates~\cite{DuanKimble2004,Reiserer2014,Welte2021}. The spin state $\ket\uparrow$ couples to the cavity mode through an optical transition, whereas $\ket\downarrow$ does not. If the spin is $\ket\downarrow$, a photon resonant with the empty cavity enters it and leaves with a phase $\pi$. If the spin is $\ket\uparrow$ and the atom--cavity coupling is strong, the coupling splits the cavity resonance away from the photon frequency, and the photon is reflected by the input mirror without this phase. For a pulse much longer than the cavity response time, every photon acquires the same conditional phase. One reflection thus applies $e^{i\pi a^\dagger a}$ to the pulse mode $a$ when the spin is $\ket\downarrow$ and leaves the light unchanged when it is $\ket\uparrow$.

For each bond the experimenter sends a coherent pulse $\ket\alpha$, with mean photon number $n=|\alpha|^2$, to reflect from the cavities at $u$ and $v$ in turn. The phases of the two reflections add, and two $\pi$ phases cancel. The light thus acquires a phase $\pi$ exactly when the bond is anti-aligned,
\begin{equation}
 U^{\rm opt}_e=\Pi^e_+\otimes I+\Pi^e_-\otimes e^{i\pi a^\dagger a}.
 \label{eq:optical-gate}
\end{equation}
Since $e^{i\pi a^\dagger a}\ket\alpha=\ket{-\alpha}$, the pulse leaves as $\ket{(-1)^{x_{uv}}\alpha}$. The spins now control the phase of the light, the reverse of Eq.~\eqref{eq:controlled-parity}, but the record map has the form of Eq.~\eqref{eq:coupling}, with records $\ket{\pm\alpha}$ and
\begin{equation}
 c=e^{-2n},\qquad s=1-e^{-4n}.
\end{equation}
A Kennedy receiver~\cite{Kennedy1973} displaces the pulse by $+\alpha$ and detects it with an on/off detector. It maps the negative record $\ket{-\alpha}$ to vacuum and clicks on the positive record with probability $s$, so it is the witness measurement of Sec.~\ref{sec:model} in optical form.

Theorem~\ref{thm:local} therefore gives an exact limit for any receiver that treats the pulses separately, and a Kennedy receiver on each pulse followed by the coverage rule attains it. The limit covers arbitrary local processing, local reference beams and classical feed-forward between pulses, including adaptive Dolinar-type detection within each pulse~\cite{Dolinar1973}. A receiver that interferes different record pulses, like the one below, is a joint receiver.

\subsection{From cancellation to one click}

On the triangle the best per-mode receiver reaches $3s^2-2s^3\simeq48n^2$. The joint receiver instead brings the three pulses to a balanced three-port at the same time and in the same spatial and temporal mode, keeps the sum mode $(a_1+a_2+a_3)/\sqrt3$, displaces it by $+\alpha/\sqrt3$ and detects it [Fig.~\ref{fig:optics}(a)]. Each alternative has two negative bonds, so its sum-mode amplitude $-\alpha/\sqrt3$ is displaced to exact vacuum. The target amplitude becomes $4\alpha/\sqrt3$. A click therefore excludes every alternative. All-up and all-down send identical pulses to the receiver, so the click cannot distinguish them, and the receiver implements the faithful projection with efficiency
\begin{equation}
 \Lin=1-e^{-16n/3},\qquad
 \Glo(C_3)=\frac{s(4-3s)}{3-2s}.
 \label{eq:triangle}
\end{equation}
Both efficiencies begin at $16n/3$, so this fixed receiver attains the optimal leading order. Its efficiency exceeds that of \emph{every} per-mode receiver by a factor that grows as $1/(9n)$ at weak light [Fig.~\ref{fig:optics}(b)]. At stronger recording, separate Kennedy receivers eventually overtake it, with the crossover near $n=0.34$. Displacement nulling and multimode interference are established tools for unambiguous optical discrimination~\cite{vanEnk2002,Sidhu2023,Guha2011}. Here they yield a gain over the exact optimum of all per-mode strategies.

The displacement requires a reference beam phase-locked to the record pulses. This reference light is not counted in $n$, which counts only the photons that interact with the spins. The ideal record map also assumes unit reflectivity for both spin states and a reflection phase difference of exactly $\pi$ between them. Finite cooperativity, unequal reflectivities, scattering and phase drift require a platform-specific assessment.

\subsection{Why work at weak light?}

Light lost during recording gives photon number a physical cost. If a fraction $\ell$ escapes between the two reflections, the escaping light has seen only one spin and partly reveals its orientation. This reduces the coherence between all-up and all-down by $e^{-2\ell n}$ per bond, and weaker pulses leak less. Uniform loss after both reflections does not damage the coherence. It leaves the conditional records pure and simply replaces $n$ by the transmitted photon number. The two target configurations then leave the same record in the environment as well, so the projection remains faithful.

Dark counts set the opposite limit. Once signal clicks become rare compared with dark clicks, clicks stop selecting the target. For $\ket+^{\otimes3}$ and a fixed nonzero dark-click probability, the squared GHZ fidelity then approaches the target weight $1/4$. Appendix~\ref{app:optics} gives the yield and fidelity of this receiver. These loss and dark-count models describe only this receiver. Section~\ref{sec:noise} compares local and joint receivers at a common fidelity under a separate probe-noise model.

%=====================================================================
\section{Targets spanning several patterns: preserving coherence}\label{sec:multi}

A target may also contain several bond patterns, as when it fixes only the parity of two distant spins. Success must then preserve coherence between spin states that leave different probe records. Consider three spins on an open chain, with probes on bonds $12$ and $23$, and take as target the end-spin parity $Z_1Z_3=+1$. Since $Z_1Z_3=(Z_1Z_2)(Z_2Z_3)$, the target contains the bond patterns $00$ and $11$. The spin state $\ket{\uparrow\uparrow\uparrow}$ leaves the probe record $\ket{a_+a_+}$, whereas $\ket{\uparrow\downarrow\uparrow}$ leaves $\ket{a_-a_-}$. A faithful projection $P_F\Pi_S\rho\Pi_S$, with $\Pi_S=\Pi_{00}+\Pi_{11}$, must keep the coherence between such states.

Reading both bond signs and accepting when they agree excludes the wrong patterns but reveals which target pattern occurred. The components $\ket{\uparrow\uparrow\uparrow}$ and $\ket{\uparrow\downarrow\uparrow}$ then end in different accepted branches, and their coherence is lost. A faithful effect must instead respond identically to the two target records,
$\bra{w_{00}}F\ket{w_{00}}=\bra{w_{11}}F\ket{w_{11}}=\bra{w_{00}}F\ket{w_{11}}$.
Equivalently, it must annihilate the difference $\ket{a_+a_+}-\ket{a_-a_-}$ as well as the two alternative records, which erases the information about which target pattern occurred. Only one direction of the four-dimensional probe space remains, and projecting onto it gives
\begin{equation}
 P^{\rm chain}_{\rm GLOBAL}=\frac{(1-c^2)^2}{2(1+c^2)}
 =\frac{s^2}{2(2-s)}.
 \label{eq:chain-alpha}
\end{equation}
For $0<c<1$ no readout that treats the two probes separately can perform this faithful projection at all, whereas the joint projection succeeds (Proposition~\ref{prop:path} below).

The chain has no loop, so joint access is needed here for a different reason than in Sec.~\ref{sec:class}. It serves to preserve coherence between target records, not to exclude alternatives more efficiently. To see how much joint access such a target requires, divide the probes into fixed disjoint blocks, and allow coherent operations inside each block and classical communication between blocks.

\begin{proposition}[Block criterion]\label{prop:path}
For the cut code of a connected graph with record overlaps in $(0,1)$, a nonzero faithful projection onto $Z_uZ_v=+1$ with a fixed block partition exists if and only if some block contains all bonds of a $u$--$v$ path. Block-separable and block-LOCC readouts are equivalent for this feasibility question.
\end{proposition}

Coherent access along a path between $u$ and $v$ is therefore the resource for this target, and some block must contain at least as many probes as the graph distance between $u$ and $v$. Appendix~\ref{app:multi} gives the general block-feasibility proof and the open-chain filter.

%=====================================================================
\section{A separation at a common output fidelity}\label{sec:noise}

When noise makes every probe record full rank, every nonzero positive effect accepts every alternative with some probability, and zero false positives become impossible. We instead fix a common minimum output fidelity and compare how often local and joint receivers accept, in one specified four-spin experiment.

The experiment uses $C_4$ with $c=4/5$ ($p=1/5$) and the spin input $\ket+^{\otimes4}$. After ideal recording, each probe is depolarized by
$X\mapsto(1-\eta)X+\eta\Tr(X)I/2$, with $\eta=0.01$, as for probes stored before readout.
For a given receiver let $f_x$ be its acceptance probability on the noisy record of pattern $x$. The yield $Y$, the probability of acceptance per round including false acceptances, and the squared fidelity $\mathcal F$ of the accepted state with $\mathrm{GHZ}_4$ are
\begin{equation}
 Y=\frac18\sum_{x\in\cC}f_x,\qquad
 \mathcal F=\frac{f_0}{\sum_{x\in\cC}f_x}.
 \label{eq:noise-metrics}
\end{equation}
The factor $1/8$ is the weight of each legal pattern in $\ket+^{\otimes4}$. Because the noise acts only on the probes after the records are written, the target block of the accepted state keeps its GHZ coherence, and $\mathcal F$ is the fraction of accepted weight in the target pattern (Appendix~\ref{app:noise}).

Now require $\mathcal F\ge0.97$. An exact dual certificate bounds the yield of every acceptance effect that is positive under partial transposition (PPT) across all seven bipartitions of the four probes~\cite{Peres1996}. Separable effects are PPT, so the bound covers every SEP and LOCC receiver~\cite{Rains2001,Cosentino2013},
\begin{equation}
 Y_{\rm local}<0.015\qquad(\mathcal F\ge0.97).
 \label{eq:Uloc}
\end{equation}
A fixed joint circuit exceeds this bound even with noisy gates. It applies single-qubit rotations and six CNOTs in depth three between probes on adjacent bonds, and accepts when all four probes read $0$. After each CNOT, a two-qubit depolarizing replacement acts with probability $g$, and all other readout operations are ideal. Both receivers face the same record noise $\eta$, but only the joint receiver is charged for gate noise. For every $0\le g\le10^{-3}$,
\begin{equation}
 \mathcal F\ge0.97,\qquad Y_{\rm joint}>0.024.
 \label{eq:circuit}
\end{equation}
The joint circuit thus delivers states of the required quality with a yield more than $60\%$ above the bound on every local strategy.

Both displayed bounds are rounded conservatively from exact certificates (Appendix~\ref{app:noise}) and hold for this input and these noise channels. Noise acting on the spins during recording, and the optical dark counts of Sec.~\ref{sec:optics}, would require separate comparisons.

%=====================================================================
\section{Discussion}\label{sec:discussion}

A local receiver can only collect certain evidence, one witness at a time, until every alternative is excluded. Its exact limit is therefore a coverage probability, and simultaneous local measurements whose outcomes are collected afterwards already attain it. A joint receiver can also accept coherent combinations of ambiguous outcomes in which the alternatives cancel. Whether such combinations exist depends on which records can occur. For pairwise parity records a loop provides exactly the needed constraint, whereas repeated records only reduce to two-state problems that local readout already solves optimally.

The relative gain is largest at weak recording, where joint readout lowers the leading power of the conditional efficiency in $s$. The local power counts the independent bond signs that must be witnessed. The joint power is set by the smallest eigenvalues of the record Gram matrix, which measure how close the legal records come to linear dependence. For pairwise parity records the joint power is at least half the local one. The optical triangle shows the gain in its simplest form. One click behind a three-pulse interferometer replaces the two witnesses that every per-mode receiver needs, and the efficiency rises from order $n^2$ to order $n$.

A target that contains several bond patterns also requires the readout to erase the information that distinguishes the target records. This can demand coherent access along a path even on a tree, where single-pattern filtering gains nothing from joint readout. With full-rank record noise, zero false positives are impossible, and receivers must instead be compared at a common output fidelity. In the specified four-spin example, a joint circuit with noisy gates still exceeds the yield bound on every local receiver.

Our scaling statements hold at fixed graph size. On a large lattice at fixed recording strength, the natural targets are parities of distant spins, and faithful projection onto them requires coherent access along a path (Sec.~\ref{sec:multi}). Reading such a parity without projecting onto it is analyzed for the same records, with overlap $c=1-p$, on the square lattice in a companion work~\cite{MiaoInPrep}. When a receiver only guesses whether two distant spins are aligned, every separable or PPT readout approaches random guessing as the separation grows for $0<p\le1-1/\sqrt2$. In terms of the witness probability this is $s\le1/2$, the bond-percolation threshold of the square lattice~\cite{Kesten1980}. As in Theorem~\ref{thm:local}, witnessed bonds can link distant spins only if they percolate~\cite{MiaoInPrep}. A joint readout instead keeps a distance-independent advantage for $p\ge0.292$, so the two readouts separate in the window $0.292\le p\le1-1/\sqrt2\simeq0.2929$. On a tree, separate measurements along a path already read the parity optimally, so loops are again what separates the readouts~\cite{MiaoInPrep}. Whether faithful projection onto distant parities shows a corresponding long-range separation in efficiency is open. So is how much of the gain survives when joint readout is limited in the number of coherently measured probes and in circuit depth. Three-probe blocks already guarantee a strict gain at fixed size, but neither the block size needed to reach the optimum nor the best efficiency at fixed block size is known in general. Other open problems are tight separable limits for filtering when some false positives are allowed, and an optical comparison at common fidelity that includes dark counts.

%=====================================================================
\begin{acknowledgments}
The author used ChatGPT (OpenAI) and Claude (Anthropic) to help build the framework of the manuscript and to refine its exposition. Claude also generated the exact certificates and the joint circuit used in Sec.~\ref{sec:noise}. The author takes responsibility for the scientific claims and the final text.
\end{acknowledgments}

%=====================================================================
\appendix

\section{Faithful filtering and general check records}\label{app:faithful}

\emph{Faithful operation.} Collect all accepting probe Kraus operators into one map $K$ that keeps their outcome register, so that $K^\dagger K=F$. Since $\|K\ket{w_x}\|^2=\bra{w_x}F\ket{w_x}$, Eq.~\eqref{eq:zfp} gives $K\ket{w_x}=0$ for $x\ne0$. Hence
\begin{equation}
 (I\otimes K)V=\Pi_0\otimes K\ket{w_0}.
\end{equation}
Tracing out the probes and the outcome register leaves $P_F\Pi_0\rho\Pi_0$. Since $K$ acts only on the probes, this proves Eq.~\eqref{eq:faithful} also for spins entangled with a reference. Conversely, an input supported on an alternative pattern $x$ is accepted with probability $\bra{w_x}F\ket{w_x}$, so faithfulness requires Eq.~\eqref{eq:zfp}.

\emph{Commuting checks.} Replace the bond parities by commuting Hermitian Pauli checks $h_e$, with their signs included, and record each check as in Eq.~\eqref{eq:coupling}, with $x_e=1$ marking the eigenvalue $-1$ of $h_e$. Let $\mathcal D$ consist of the binary dependencies satisfying
$\prod_e h_e^{\lambda_e}=(-1)^{\epsilon(\lambda)}I$.
The legal patterns obey $\lambda\cdot x=\epsilon(\lambda)$ for every $\lambda\in\mathcal D$ and form an affine coset of $\mathcal D^\perp$. They form the linear code $\cC=\mathcal D^\perp$ when the check group excludes $-I$. Choosing any legal target $x_*$ and exchanging the two record labels on its negative coordinates maps that coset to $\mathcal D^\perp$ and the target to $0$. This separate relabeling on each probe preserves all overlaps and measurement classes. The general linear-code results therefore apply to any such legal single-pattern target.

\section{The local bound}\label{app:local}

We prove the bound for the larger class in which only the acceptance effect is separable. Writing its optimum as $P_{F\text{-sep}}$ gives
\begin{equation}
 \Loc\le\Sep\le P_{F\text{-sep}},\qquad \Sep\le\Glo.
 \label{eq:chain}
\end{equation}
\emph{Proof of Theorem~\ref{thm:local}.} Write $\ket{w_S}$ for the record with $\ket{a_-}$ on $S$ and $\ket{a_+}$ elsewhere, and $\ket{w_\varnothing}$ for the target. By the compression described below, it suffices to work on the two-dimensional record planes. A separable $F$ is a finite sum $\sum_\ell\bigotimes_eA^{(\ell)}_e$ with $A^{(\ell)}_e\ge0$. Dropping the terms with zero target expectation leaves an operator $F_+$ with $0\le F_+\le F\le I$ and the same efficiency. Each factor of $F_+$ has $\bra{a_+}A^{(\ell)}_e\ket{a_+}>0$.

For $S\in\cS$ every term has nonnegative expectation in $\ket{w_S}$ and the sum vanishes, so each term vanishes. The factors outside $S$ are positive on $\ket{a_+}$, so some $e\in S$ has $\bra{a_-}A^{(\ell)}_e\ket{a_-}=0$. Positivity then gives $A^{(\ell)}_e\ket{a_-}=0$, which places $\ran A^{(\ell)}_e$ inside $\ran P_e$. These probes form a set $T_\ell$ that meets every $S\in\cS$. Expanding the other factors in the basis $\{P_e,I-P_e\}$, every basis string in the range of the term has a witnessed set containing $T_\ell$, which again meets every $S$. The range of $F_+$ thus lies in that of the projector $Q_\cS$ onto the covering strings, and $F_+\le I$ gives $F_+\le Q_\cS$. On $\ket{w_\varnothing}=\bigotimes_e\ket{a_+}$ the outcomes are independent with $\Pr(P_e)=s_e$, so $\bra{w_\varnothing}F\ket{w_\varnothing}\le\bra{w_\varnothing}Q_\cS\ket{w_\varnothing}=H_\cS$.

Conversely, every string accepted by $Q_\cS$ has a $P$ outcome on some $e\in S$, where $\ket{w_S}$ carries $\ket{a_-}$, so $Q_\cS$ has zero false positives. Both $Q_\cS$ and $I-Q_\cS$ are sums of orthogonal product projectors, so they form a one-round LOCC measurement. Together with Eq.~\eqref{eq:chain}, $H_\cS\le\Loc\le\Sep\le P_{F\text{-sep}}\le H_\cS$. Closedness of the separable cone covers limits of LOCC protocols~\cite{Chitambar2014}. $\square$

For probes of larger dimension, let $\mathcal K_e$ be the span of their two records. Compression by $\Pi_\mathcal K=\bigotimes_e\Pi_{\mathcal K_e}$ preserves separability, the bound $0\le F\le I$, and every record expectation. The argument on the two-dimensional record planes thus also covers optical modes, local ancillas and classical feed-forward. Unequal overlaps enter only through the independent witness probabilities $s_e$.

For a single alternative of total overlap $\gamma=\prod_{e\in S}c_e$, the coverage probability is $1-\gamma^2$. It is the one-sided limit in which only the target is ever identified, a special case of unambiguous discrimination with unequal priors~\cite{JaegerShimony1995}. Equal-prior unambiguous discrimination of the two states instead succeeds with probability $1-\gamma$~\cite{Ivanovic1987,Dieks1988,Peres1988}, and local operations attain this value for product states~\cite{ChenYang2001,JiCaoYing2005}.

\section{The joint optimum}\label{app:global}

The geometric pure-state filtering construction~\cite{Bergou2003,BergouHerzogHillery2005} projects the target onto the complement of the alternative span. Within the record span, the dual vector~\cite{Chefles1998}
\[
 \ket\xi=\sum_x(\Gamma^{-1})_{x0}\ket{w_x}
\]
satisfies $\ip{w_x}{\xi}=\delta_{x0}$, with $\Gamma_{xy}=\prod_e c_e^{(x+y)_e}$. It spans that complement, and its normalized projector has target efficiency $1/\ip{\xi}{\xi}=1/(\Gamma^{-1})_{00}$.

Because $\Gamma_{xy}$ depends only on $x+y$, the Gram matrix is diagonalized by the characters of the group $\cC$, as for any group-covariant state set~\cite{CheflesBarnett1998,EldarForney2001}. For $x=u^{\mathsf T}\G$ the character labeled by $\sigma\in\F^k$ is $\chi_\sigma(x)=(-1)^{\sigma\cdot u}$. The same structure underlies the code setting of Ref.~\cite{Mandal2026}. Substituting $c_e^{x_e}=\mathbb E[(-1)^{x_eZ_e}]$, where $\Pr(Z_e=1)=(1-c_e)/2$, identifies the eigenvalues of $\Gamma$ as
\begin{equation}
 \lambda_\sigma=2^k\Pr[\G Z=\sigma]=2^k\pi(\sigma).
\end{equation}
The normalized eigenvector $2^{-k/2}\chi_\sigma$ has target component $2^{-k/2}$. Thus
\begin{equation}
 \Glo=\frac{1}{(\Gamma^{-1})_{00}}
 =4^k\Bigl[\sum_{\sigma\in\F^k}\pi(\sigma)^{-1}\Bigr]^{-1},
 \label{eq:coset}
\end{equation}
which is the harmonic mean of the $2^k$ eigenvalues $\lambda_\sigma$. For $0<c_e<1$ every $\pi(\sigma)$ is positive, so the legal record states are linearly independent, and the pattern constraints restrict only which states occur.

For the $C_4$ curves in Fig.~\ref{fig:class}, the Gram eigenvalues are $s^2-8s+8$, $s(2-s)$ (fourfold) and $s^2$ (threefold). Substitution gives
\begin{align}
 \Glo(C_4)&=\frac{s^2(2-s)(s^2-8s+8)}{6-5s},\nonumber\\
 \Loc(C_4)&=4s^3-3s^4.
 \label{eq:C4}
\end{align}
Their difference is $16s^2(1-s)^3/(6-5s)>0$ on $0<s<1$.

\section{Classification: full proof}\label{app:class}

Zero columns correspond to probes that always hold $\ket{a_+}$. They carry no information and can be discarded, so assume all columns are nonzero.

(ii)$\Leftrightarrow$(iii). Two nonzero columns are dependent if and only if they are equal. Suppose every circuit has at most two elements. A dependency among distinct nonzero columns would contain a circuit with at least three elements, so the classes $B_1,\dots,B_r$ of equal columns have linearly independent representatives, and $r=k$. A codeword is then constant on each class, and the classes take independent values, so $\cC=\bigoplus_j\mathrm{Rep}(B_j)$, where the repetition code $\mathrm{Rep}(B_j)$ consists of the all-zero and all-one words on $B_j$. Conversely, the column matroid of such a direct sum has only two-element circuits~\cite{Oxley2011}.

\emph{No gap for repetition blocks.} When $\cC$ is the direct sum of repetition codes on the blocks $B_j$, each block is a two-state problem with records $\bigotimes_{e\in B_j}\ket{a_\pm}$ and overlap $\gamma_j=\prod_{e\in B_j}c_e$. The Gram matrix factorizes over blocks, so $\Glo=\prod_j(1-\gamma_j^2)$. A set $T$ covers all codeword supports exactly when it meets every block, so $H_\cS=\prod_j(1-\gamma_j^2)$ as well.

(ii)$\Rightarrow$ gap. Let $D$ be a circuit with $|D|\ge3$, $J=\{d_0,d_1,d_2\}\subseteq D$, and extend $D\setminus\{d_0\}$ to a basis $\mathcal B$. Put $W'=\mathcal B\setminus\{d_1,d_2\}$ and $R=[m]\setminus(\mathcal B\cup\{d_0\})$, so that $[m]=W'\cup R\cup J$ disjointly, and $\Pi_{\rm br}=\bigotimes_{e\in W'}P_e\otimes\bigotimes_{e\in R}(I-P_e)$. The protocol measures every probe outside $J$ locally in $\{P_e,I-P_e\}$. Only in the branch $\Pi_{\rm br}$, in which the witnesses outside $J$ are exactly the probes in $W'$, does it measure the three probes of $J$ jointly.
(a) \emph{Survivors.} $\Pi_{\rm br}$ annihilates every record with $x_e=1$ for some $e\in W'$. On $R$, $I-P_e$ maps both records to nonzero multiples of $\ket{a^{(e)}_-}$ and imposes no constraint. If $x|_{W'}=0$, then $x|_{\mathcal B}$ is supported in $\{d_1,d_2\}$, and since $D\setminus\{d_0,d_1,d_2\}\subseteq W'$, the circuit gives $x_{d_0}=x_{d_1}+x_{d_2}$. The survivors restricted to $J$ are therefore $000$ (only $x=0$), $110$, $101$ and $011$, the legal patterns of a triangle.
(b) \emph{Local rule in the branch.} Let $W_J\subseteq J$ be the witnessed probes of $J$. The set $W'\cup W_J$ has rank $k$ if and only if $|W_J|\ge2$, so $Q_\cS(\Pi_{\rm br}\otimes I_J)=\Pi_{\rm br}\otimes Q_\triangle$, with $Q_\triangle$ the projector onto strings with at least two witnesses on $J$.
(c) \emph{Extra direction.} With $\ket{a^{(i)}_+}=c_i\ket{a^{(i)}_-}+t_i\ket{a^{(i)\perp}_-}$ set
\begin{equation}
 \ket{\tilde v}=\ket{a_-a_-a_-}-\sum_{i\in J}\frac{c_i}{t_i}\ket{a^\perp_-}_i\ket{a_-a_-}_{J\setminus i}.
\end{equation}
The survivor with $\ket{a_+}$ on $i$ has overlap $c_i-c_i=0$ with $\ket{\tilde v}$. Every term of $\ket{\tilde v}$ has at most one witness, so $Q_\triangle\ket{\tilde v}=0$. The target overlap is $-2c_{d_0}c_{d_1}c_{d_2}$. With $\ket v=\ket{\tilde v}/\|\tilde v\|$, the effect $F=Q_\cS+\Pi_{\rm br}\otimes\ket v\bra v$ is a sum of orthogonal projectors that annihilates every alternative, and
\begin{align}
 \bra{w_0}F\ket{w_0}&=H_\cS+\prod_{e\in W'}s_e\prod_{e\in R}(1-s_e)\nonumber\\
 &\quad\times\frac{4c_{d_0}^2c_{d_1}^2c_{d_2}^2}{1+\sum_{i\in J}c_i^2/t_i^2}>H_\cS .
\end{align}
For identical overlaps the increment is $4s^{k-1}(1-s)^{m-k+2}/(3-2s)$. The protocol uses only one joint measurement, on three probes, and measures $J$ locally in every other branch.

\emph{Graph version.} For a connected simple graph with a cycle $C$, choose a connected spanning subgraph $H\supseteq C$ with $N$ edges whose only cycle is $C$, three edges $J$ of $C$, and $T=E(H)\setminus J$. In the branch in which exactly $T$ witnesses, every surviving alternative restricts to a two-edge cut of the triangle obtained by contracting the three components of $T$, and the protocol has efficiency
\begin{equation}
 P_3=\mathcal R_G(s)+\frac{4s^{N-2}(1-s)^{m-N+3}}{3-2s},
\end{equation}
independent of the choices of $C$, $H$ and $J$. On the bare triangle this construction is the optimal filter.

\section{Weak-recording exponents}\label{app:weak}

For the constants in Proposition~\ref{prop:exponents}, let $r_\sigma=\min_{\G z=\sigma}|z|$, let $\nu_\sigma$ count its minimizers, let $b$ count the column bases, that is, the sets of $k$ linearly independent columns, and put $\Delta=\sum_{\sigma:r_\sigma=\rc}\nu_\sigma^{-1}$. Then $A_\cC=4^{k-\rc}/\Delta$.

\emph{Proof.} In the local coverage probability, the smallest full-rank witness sets are the $k$-element bases. Their contributions give $\Loc=b s^k[1+O_\cC(s)]$. For the joint filter,
\begin{equation}
 \pi(\sigma)=\sum_{\G z=\sigma}q^{|z|}(1-q)^{m-|z|}
 =\nu_\sigma q^{r_\sigma}[1+O_\cC(q)].
\end{equation}
The sectors with the largest $r_\sigma=\rc$ dominate, so $\sum_\sigma\pi(\sigma)^{-1}=\Delta q^{-\rc}[1+O_\cC(q)]$. Using $q=s/4+O(s^2)$ in Eq.~\eqref{eq:coset} proves the joint expansion.

Every syndrome is a sum of at most $k$ columns from a fixed basis, so $\rc\le k$. Now let a circuit $D$ have at least three elements, choose $d_0\in D$ and a basis $\mathcal B\supseteq D\setminus\{d_0\}$. Expanded in $\mathcal B$, only the syndrome $\sum_{e\in\mathcal B}\mathsf g_e$ can require $k$ columns. Since the circuit gives $\sum_{e\in D\setminus\{d_0\}}\mathsf g_e=\mathsf g_{d_0}$, this syndrome also equals
$\mathsf g_{d_0}+\sum_{e\in\mathcal B\setminus D}\mathsf g_e$,
a sum of $k-|D|+2\le k-1$ columns. Thus $\rc<k$. For a direct sum of repetition codes, each syndrome is a unique sum of the $k$ class representatives and needs one column from each class it uses, so $\rc=k$. $\square$

For graph records, each syndrome corresponds to an even set $T$ of vertices, and its minimal weight is the size of a minimum $T$-join, the standard cycle-code correspondence~\cite{SoleZaslavsky1993}. A $T$-join meets every vertex of $T$, so it has at least $|T|/2$ edges. Choosing $T$ to contain all vertices when $N$ is even, or all but one when $N$ is odd, gives $\rc\ge\lfloor N/2\rfloor$ and hence Eq.~\eqref{eq:halving}. For a single cycle of length $m$, the dependency code consists of the zero vector and the all-ones vector. A binary vector of weight $w$ is therefore at distance $\min(w,m-w)$ from this code, giving $\rc=\lfloor m/2\rfloor$. Odd cycles have $\rc/(N-1)=1/2$, whereas finite even cycles lie above one half.

\section{Loss and dark counts in the optical receiver}\label{app:optics}

Loss after writing an ideal record maps
$\ket{w_x}\mapsto\ket{\tilde w_x}\ket{e_x}_E$.
If the readout annihilates every attenuated alternative $\ket{\tilde w_x}$, the accepted state carries the common target environment record $\ket{e_0}$ and remains proportional to $\Pi_0\rho\Pi_0$. A uniform transmission $\mathcal T$ replaces $n$ by $\mathcal Tn$. Nonuniform transmission requires rebalancing the sum mode.

Loss between the two reflections has a different effect. The escaping light has interacted with only the first spin and distinguishes its two orientations. If a fraction $\ell$ is lost, the overlap of the two environment states is
\begin{equation}
 \ip{-\sqrt\ell\alpha}{\sqrt\ell\alpha}=e^{-2\ell n}.
\end{equation}
This factor multiplies the coherence between all-up and all-down for each such bond.

For a signal-independent dark-click probability $d$, the target click probability is $f_0=1-(1-d)e^{-16\mathcal Tn/3}$, and each of the three alternatives clicks with probability $d$. On input $\ket+^{\otimes3}$,
\begin{equation}
 Y=\frac{f_0+3d}{4},\qquad \mathcal F=\frac{f_0}{f_0+3d}.
 \label{eq:dark}
\end{equation}
Here $Y$ includes false acceptances and $\mathcal F$ is squared GHZ fidelity. At fixed $d>0$ and $n\to0$, the clicks carry no pattern information and $\mathcal F\to1/4$.

\section{The noisy comparison}\label{app:noise}

\subsection{Accepted spin state}

Let ideal recording be followed by a completely positive trace-preserving channel $\cN$ acting only on probes. Set $f_x=\Tr[F\cN(\ket{w_x}\bra{w_x})]$. The unnormalized accepted spin state has the form
\begin{equation}
 \mathcal I=\sum_{x,y}M_{xy}\Pi_x\rho\Pi_y,\quad
 M_{xy}=\Tr[F\cN(\ket{w_x}\bra{w_y})].
 \label{eq:schur}
\end{equation}
\begin{lemma}[Probe-side noise]\label{lem:fid}
For $\omega_0=\Tr(\Pi_0\rho)>0$ and a nonzero acceptance probability,
\begin{equation}
 \Pi_0\mathcal I\Pi_0=f_0\Pi_0\rho\Pi_0,\qquad
 \Tr\mathcal I=\sum_x f_x\Tr(\Pi_x\rho).
\end{equation}
The squared fidelity of $\mathcal I/\Tr\mathcal I$ with $\Pi_0\rho\Pi_0/\omega_0$ is $f_0\omega_0/\Tr\mathcal I$.
\end{lemma}
\emph{Proof.} The two identities follow from Eq.~\eqref{eq:schur}. Writing $\tau=\Pi_0\rho\Pi_0/\omega_0$ and $\kappa=f_0\omega_0/\Tr\mathcal I$, the target block of the normalized accepted state is $\kappa\tau$. Since $\sqrt\tau$ is supported on $\Pi_0$, the squared fidelity is
$(\Tr\sqrt{\kappa\tau^2})^2=\kappa$. $\square$

For $\ket+^{\otimes4}$ on $C_4$, all eight legal patterns have weight $1/8$, and the lemma gives Eq.~\eqref{eq:noise-metrics}.

\subsection{Local upper bound and joint circuit}

With the noisy records $\rho_x$ of Sec.~\ref{sec:noise}, put $A=\frac18\sum_x\rho_x$, $B=\rho_0/8$, and $C=B-0.97A$. Then $Y=\Tr(FA)$ and the requirement $\mathcal F\ge0.97$ is $\Tr(FC)\ge0$. A dual certificate consists of $\lambda\ge0$ and positive semidefinite matrices $Z$ and $W_\beta$ such that
\begin{equation}
 K=Z-A-\lambda C-\sum_\beta W_\beta^{T_\beta}\ge0,
\end{equation}
where $\beta$ runs over the seven probe bipartitions and $T_\beta$ is the corresponding partial transpose. Let $0\le F\le I$ with $F^{T_\beta}\ge0$ and $\Tr(FC)\ge0$. Then $\Tr(FW_\beta^{T_\beta})=\Tr(F^{T_\beta}W_\beta)\ge0$ and $\lambda\Tr(FC)\ge0$, so $\Tr(FK)\ge0$ gives $Y\le\Tr(FZ)\le\Tr Z$. A rational certificate, checked in exact arithmetic, has $\Tr Z<0.015$. Since separable effects are PPT, it bounds every separable acceptance effect and hence every LOCC protocol.

The joint circuit acts on the probes in the basis of Eq.~\eqref{eq:probe-preparation}. Label the probes $0,1,2,3$ on the bonds $(12,23,34,41)$ of $C_4$, and let $R_j(\theta)=e^{-i\theta\sigma^y_j/2}$ be a real rotation of probe $j$. In time order, a two-CNOT block on probes $(a,b)$ applies $R_a(\theta_1)R_b(\theta_2)$, $\mathrm{CNOT}_{a\to b}$, $R_a(\theta_3)R_b(\theta_4)$, $\mathrm{CNOT}_{b\to a}$ and $R_a(\theta_5)R_b(\theta_6)$. The circuit applies this block to the pairs $(a,b)=(0,1)$ and $(3,2)$ in parallel, then $\mathrm{CNOT}_{0\to3}$ and $\mathrm{CNOT}_{1\to2}$ in parallel, then $R_0(\theta_7)R_1(\theta_8)$, and accepts when all four probes read $0$. In radians the angles are
\begin{equation}
\begin{aligned}
 \theta_1&=-0.633, &\theta_2&=-1.989, &\theta_3&=2.423,\\
 \theta_4&=-0.886, &\theta_5&=-2.401, &\theta_6&=-1.694,\\
 \theta_7&=0.643, &\theta_8&=3.113.
\end{aligned}
\label{eq:circuit-angles}
\end{equation}
The certified circuit uses nearby angles with rational $\tan(\theta/4)$, so that all its matrix elements are rational. Rounding them to the three decimals shown changes $Y$ and $\mathcal F$ by less than $10^{-5}$. With $\eta=0.01$ and ideal gates, the circuit gives $Y=0.0248$ and $\mathcal F=0.983$. Following each CNOT on probes $a,b$, the gate-noise channel is
\begin{equation}
 X\mapsto(1-g)X+g\,\frac{I_{ab}}4\otimes\Tr_{ab}X.
\end{equation}
All other operations are ideal. The pattern acceptance probabilities of the noisy circuit are then polynomials in $g$. Interval bounds on these polynomials prove $\mathcal F\ge0.97$ and $Y>0.024$ throughout $0\le g\le10^{-3}$. The local receiver is charged only for the common record noise.

\section{Multi-pattern targets}\label{app:multi}

Let $\emptyset\neq S\subseteq\cC$ be the target with $\Pi_S=\sum_{x\in S}\Pi_x$, and $M_{xy}=\bra{w_y}F\ket{w_x}$. Inserting $\rho=\ket{\phi}\bra{\phi'}$ with $\phi\in\ran\Pi_x$, $\phi'\in\ran\Pi_y$ shows that the accepted branch equals $P_F\Pi_S\rho\Pi_S$ for all inputs if and only if $M=P_F\mathbf 1_S\mathbf 1_S^{\mathsf T}$, that is, $F$ annihilates every record outside $S$ and every difference of two target records. For $|S|=1$ this is Eq.~\eqref{eq:zfp}.

\emph{Open chain.} Choose local probe bases with $\ket{a_+}=\ket0$ and $\ket{a_-}=c\ket0+t\ket1$, which preserves all record overlaps. The vector $\ket f=t^2\ket{00}-ct\ket{01}-ct\ket{10}+(1+c^2)\ket{11}$ is orthogonal to $\ket{a_+a_-}$, $\ket{a_-a_+}$ and $\ket{a_+a_+}-\ket{a_-a_-}$, which are linearly independent. An admissible $F$ must annihilate these three vectors, so $F\le\ket f\bra f/\ip ff$. With $\ip ff=2(1+c^2)$ and $\ip f{a_+a_+}=\ip f{a_-a_-}=t^2$, this gives Eq.~\eqref{eq:chain-alpha}, attained by the projector onto $\ket f$.

For a partition $\mathsf P$ of the probes into blocks, let $S|_B$ be the local patterns that $S$ takes on block $B$ and $\mathrm{Sat}_{\mathsf P}(S)=\cC\cap\prod_B S|_B$.

\begin{proposition}[Block feasibility]\label{prop:block}
A block-separable effect implements $P_F\Pi_S\cdot\Pi_S$ with $P_F>0$ if and only if a block-LOCC protocol does, if and only if $S=\mathrm{Sat}_{\mathsf P}(S)$.
\end{proposition}
\emph{Proof.} Write $F=\sum_\ell\bigotimes_BA_{B,\ell}$ and $v_\ell(x)=\bigotimes_B\sqrt{A_{B,\ell}}\ket{w_{x_B}}$, so $M_{xy}=\sum_\ell\ip{v_\ell(y)}{v_\ell(x)}$. For $x\notin S$, $M_{xx}=0$ gives $v_\ell(x)=0$. For $x,y\in S$, $\sum_\ell\|v_\ell(x)-v_\ell(y)\|^2=M_{xx}+M_{yy}-2\,\mathrm{Re}\,M_{xy}=0$, so each $v_\ell$ is constant on $S$, and some $\ell$ has a nonzero constant. Every block factor of this term is then nonzero on each pattern in $S|_B$, so $v_\ell(x)\ne0$ for all $x\in\mathrm{Sat}_{\mathsf P}(S)$, which forces $\mathrm{Sat}_{\mathsf P}(S)\subseteq S$. Conversely, if $S=\mathrm{Sat}_{\mathsf P}(S)$, each block measures the projector onto $\ket{f_B}=\sum_{z\in S|_B}\ket{\tilde w_z}$, with $\ket{\tilde w_z}$ dual to the linearly independent block records. The run is accepted when all blocks succeed, which happens with the same amplitude for every target pattern and never for a pattern outside $\mathrm{Sat}_{\mathsf P}(S)$. $\square$

For $S=\{Z_uZ_v=+1\}$ on a cut code, suppose first that a block contains a $u$--$v$ path. The product of the signs along the path is $Z_uZ_v$, so every pattern in $\mathrm{Sat}_{\mathsf P}(S)$ is a target pattern and $\mathrm{Sat}_{\mathsf P}(S)=S$. If no block contains such a path, take a nontarget pattern, the cut of a set $A$ with $u\in A$ and $v\notin A$. For each block $B$, flipping the component of $u$ in the subgraph formed by the bonds of $B$ leaves the signs on $B$ unchanged and turns this cut into a target pattern. The nontarget pattern thus agrees on every block with a target pattern, so $\mathrm{Sat}_{\mathsf P}(S)\ne S$. Together with Proposition~\ref{prop:block}, this proves Proposition~\ref{prop:path}. On the open chain with one probe per block, $\mathrm{Sat}_{\mathsf P}(S)$ also contains the odd patterns $01$ and $10$.

\bibliography{references}

\end{document}